\documentclass[final,authoryear,5p]{elsarticle}

\usepackage{epsfig}

\usepackage{amssymb}
\usepackage{lineno}
\usepackage{natbib}

\def\apj{{\it Astrophys. J.} }
\def\apjl{{\it Astrophys. J. Letters} }
\def\apjs{{\it Astrophys. J. Suppl.} }

\def\pasp{{\it Publ. Astron. Soc. Pac.} }
\def\pasj{{\it Publ. Astron. Soc. Japan} }
\def\procspie{{\it Proc. SPIE} }

\usepackage[ps2pdf,%
a4paper=true,%
breaklinks=true,%
colorlinks=true,%
pdfauthor={Doi et al.},%
pdftitle={Template for manuscripts in Advances in Space Research}%
]{hyperref}

\journal{Advances in Space Research}

\begin{document}

\begin{frontmatter}



\title{A Balloon-Borne Very Long Baseline Interferometry Experiment in the Stratosphere: Systems Design and Developments}
\tnotetext[footnote1]{Received 28 May 2018, Revised 20 August 2018, Accepted 14 September 2018}



\author[1,2]{Akihiro Doi\corref{cor}}
\ead{akihiro.doi@vsop.isas.jaxa.jp}
\author[3]{Yusuke Kono\corref{cor}}
\ead{kono.yusuke@nao.ac.jp}
\author[4]{Kimihiro Kimura}
\author[2,1]{Satomi Nakahara}
\author[5]{Tomoaki Oyama}
\author[4]{Nozomi Okada}
\author[1]{Yasutaka Satou}
\author[5]{Kazuyoshi Yamashita}
\author[3]{Naoko Matsumoto}
\author[1]{Mitsuhisa Baba}
\author[6]{Daisuke Yasuda}
\author[3]{Shunsaku Suzuki}
\author[1]{Yutaka Hasegawa}
\author[5]{Mareki Honma}
\author[7]{Hiroaki Tanaka}
\author[1,8]{Kosei Ishimura}
\author[1]{Yasuhiro Murata}
\author[9]{Reiho Shimomukai}
\author[10]{Tomohiro Tachi}
\author[11]{Kazuya Saito}
\author[12]{Naohiko Watanabe}
\author[1]{Nobutaka Bando}
\author[5]{Osamu Kameya}
\author[13]{Yoshinori Yonekura}
\author[14]{Mamoru Sekido}
\author[15]{Yoshiyuki Inoue}
\author[16]{Hikaru Sakamoto}
\author[6]{Nozomu Kogiso}
\author[17]{Yasuhiro Shoji}
\author[4]{Hideo Ogawa}
\author[18]{Kenta Fujisawa}
\author[1]{Masanao Narita}
\author[19]{Hiroshi Shibai}
\author[1]{Hideyuki Fuke}
\author[9,1]{Kenta Uehara}
\author[20]{Shoko Koyama}

\address[1]{The Institute of Space and Astronautical Science, Japan Aerospace Exploration Agency, 3-1-1 Yoshinodai, Chuou-ku, Sagamihara, Kanagawa 252-5210, Japan}
\address[2]{Department of Space and Astronautical Science, The Graduate University for Advanced Studies, 3-1-1 Yoshinodai, Chuou-ku, Sagamihara, Kanagawa 252-5210, Japan}
\address[3]{National Astronomical Observatory of Japan, 2-21-1 Osawa, Mitaka, Tokyo 181-8588, Japan}
\address[4]{Department of Physical Science, Graduate School of Science, Osaka Prefecture University, 1-1 Gakuen-cho, Naka-ku, Sakai, Osaka 599-8531, Japan}
\address[5]{Mizusawa VLBI Observatory, National Astronomical Observatory of Japan, 2-12 Hoshigaoka, Mizusawa, Oshu, Iwate, 023-0861, Japan}
\address[6]{Department of Aerospace Engineering, Graduate School of Engineering, Osaka Prefecture University, 1-1 Gakuen-cho, Naka-ku, Sakai, Osaka 599-8531, Japan}
\address[7]{Department of Aerospace Engineering, National Defense Academy of Japan, 1-10-20 Hishirimizu, Yokosuka Kanagawa 239-8686, Japan}
\address[8]{Department of Modern Mechanical Engineering, School of Creative Science and Engineering, Waseda University, 3-4-1 Okubo, Shinjuku-ku, Tokyo 169-8555, Japan}
\address[9]{Department of Astronomy, Graduate School of Science, The University of Tokyo, 7-3-1 Hongo, Bunkyo-ku, Tokyo 113-0033, Japan}
\address[10]{Department of General Systems Studies, Graduate School of Arts and Sciences, The University of Tokyo, 3-8-1 Komaba, Meguro-ku, Tokyo, 153-8902, Japan}
\address[11]{Graduate School of Information Science and Technology, The University of Tokyo, 7-3-1 Hongo, Bunkyo-ku, Tokyo 113-8656, Japan}
\address[12]{National Institute of Technology, Gifu College, 2236-2 Kamimakuwa, Motosu-city, Gifu 501-0495, Japan }
\address[13]{Center for Astronomy, Ibaraki University, 2-1-1 Bunkyo, Mito, Ibaraki 310-8512, Japan}
\address[14]{Kashima Space Technology Center, National Institute of Information and Communications Technology, 893-1 Hirai, Kashima, Ibaraki 314-8501, Japan}
\address[15]{Interdisciplinary Theoretical \& Mathematical Science Program (iTHEMS), RIKEN, 2-1 Hirosawa, Wako, Saitama 351-0198, Japan}
\address[16]{Department of Mechanical Engineering, Tokyo Institute of Technology, 2-12-1 Ookayama, Meguro-ku, Tokyo 152-8552, Japan }
\address[17]{Graduate School of Engineering, Osaka University, 2-1 Yamadaoka, Suita, Osaka 565-0871, Japan}
\address[18]{The Research Institute for Time Studies, Yamaguchi University, 1677-1 Yoshida, Yamaguchi, Yamaguchi 753-8511, Japan}
\address[19]{Department of Earth and Space Science, Graduate School of Science, Osaka University, 1-1, Machikaneyamacho, Toyonaka, Osaka 560-0043, Japan}
\address[20]{Institute of Astronomy \& Astrophysics, Academia Sinica, P.O. Box 23-141, Taipei 10617, Taiwan}

\cortext[cor]{Corresponding authors}

\begin{abstract}
The balloon-borne very long baseline interferometry (VLBI) experiment is a technical feasibility study for performing radio interferometry in the stratosphere. The flight model has been developed.  A balloon-borne VLBI station will be launched to establish interferometric fringes with ground-based VLBI stations distributed over the Japanese islands at an observing frequency of approximately 20~GHz as the first step. This paper describes the system design and development of a series of observing instruments and bus systems. In addition to the advantages of avoiding the atmospheric effects of absorption and fluctuation in high frequency radio observation, 
the mobility of a station can improve the sampling coverage (``$uv$-coverage'') by increasing the number of baselines by the number of ground-based counterparts for each observation day.  This benefit cannot be obtained with conventional arrays that solely comprise ground-based stations.  The balloon-borne VLBI can contribute to a future progress of research fields such as black holes by direct imaging.  
\end{abstract}


\end{frontmatter}

\parindent=0.5 cm



\section{Introduction}\label{section:introduction}
Cosmic radio waves are degraded during their propagation through Earth's atmosphere because of atmospheric attenuation that reduces the radio signals, atmospheric radiation that increases the effective system noise of a receiving telescope, and atmospheric fluctuation in the time domain that degrades radio wave coherence.  
The column density fluctuation of water vapor in the tropospheric flow is the primary cause of phase fluctuation in the interferometry response of radio telescopes. The Atacama Large Millimeter/submillimeter Array~(ALMA) on Chajnantor plateau, Chile is located at an altitude of approximately 5,000~m, where the amount of precipitable water vapor is typically 1.0~mm and falls below 0.5~mm up to 25\% of the time\footnote{https://almascience.eso.org/about-alma/atmosphere-model}. ALMA has interferometric baselines of up to 16~km for observations at $\sim70$--$960$~GHz. Intercontinental baselines using very long baseline interferometry~(VLBI) techniques have been established at 230~GHz ($\lambda1.3$~mm) by the Event Horizon Telescope project \citep{Doeleman:2009}. The primary aim of the project is to image black hole shadows, although observing frequencies of higher than $300$~GHz (submillimeter wave) might be required for clear imaging to avoid opacity effects and interstellar scattering through the atmospheres (accreting material) surrounding black holes \citep{Falcke:2000b}. On the ground, submillimeter VLBIs are potentially available at a limited number of telescopes on sites in particularly good conditions at high altitudes.

The VLBI Space Observatory Programme ({\it VSOP}; \citealt{Hirabayashi:1998}) and {\it RadioAstron} \citep{Kardashev:2013} have established space VLBI missions in which a single space-based radio telescope makes baselines to ground-based telescopes at $\leqq22$~GHz. The purpose of these missions is not to reduce atmospheric effects but to expand baseline lengths beyond the Earth's diameter. Although submilllimeter/tera-hertz (THz) interferometry can be potentially available between space--space baselines through the use of multiple satellites, it would be expensive. Scientific balloons can typically reach stratospheric altitudes of $\sim20$--$40$~km, which are above 99.9\% of the atmosphere's water content, and balloon-borne radio telescopes realize observing sites that are essentially equivalent to space in terms of radio observations.

Many balloon-borne single-dish submillimeter telescope missions have been launched to observe the cosmic microwave background. Several such missions--- Archeops \citep{Benoicirct:2002}, BOOMERanG \citep{Crill:2003}, and EBEX \citep{Reichborn-Kjennerud:2010}---used off-axis 1.3--1.5-m Gregorian telescopes to scan a wide field using a bolometer array located in the focal plane.  
Observations at high angular resolutions were made by PRONAOS \citep{Serra:2002} and BLAST \citep{Devlin:2004}, two balloon-borne experiments employing 2~m class, on-axis Cassegrain-type telescopes.
Thus, balloon-borne submillimeter radio telescopes have been technically established.  Such a platform could potentially serve as a VLBI station if it were equipped with a VLBI backend system.
However, to date, no individual interferometric element has been launched into the stratosphere. Note that position determination to an accuracy of $\lesssim1/20$ in terms of observing wavelength for each element is crucial for accumulating radio signal to establish interferometric fringes.  This requirement represents a primary technical challenge for maintaining the phase stability of stratospheric VLBI using drifting balloons, which experience pendulum motion and string vibration between the balloon and the gondola. The Antarctic Impulsive Transient Antenna (ANITA) is a balloon-borne radio interferometer that is used to investigate radio pulses originating from the interactions of cosmic ray air showers based on correlations between antenna elements located on a single gondola platform \citep{Anita-Collaboration:2009}. Far-infrared interferometry experiments, the Far-Infrared Interferometric Experiment (FITE; \citealt{Shibai:2010}) and the Balloon Experimental Twin Telescope for Infrared Interferometry (BETTII; \citealt{Rinehart:2014}), will employ a meter class baseline between two individual mirror elements located on the same gondola. Overall, maintaining a correct working relationship between individual interferometry elements in the stratosphere at submillimeter/THz regimes represents a significant technical challenge. The dynamical fluctuations in a balloon-borne platform can also affect the phase stability of the frequency standard clock, which is a critical component of VLBI system.  Furthermore, maintaining the pointing stability required to track a target source demands careful treatment. Individual radio interferometric elements generally comprise a single heterodyne receiver on a high gain antenna with a diameter of at least a few meters, thus resulting in a beam width of the order of an arcminute or less, then making absolute pointing accuracy (and not just stability) is required.  For the BLAST telescope, a pointing accuracy of less than $5$~arcsec r.m.s.~has been achieved in raster mapping \citep{Pascale:2008}; however, the dynamical fluctuation of the gondola platform necessitated post-flight pointing reconstruction.

In this paper, we present the configuration of observing systems and gondola bus systems on the flight model of a balloon-borne VLBI station that is to be used in a technical feasibility study for future stratospheric interferometric missions. The flight model was developed at the Japan Aerospace Exploration Agency~(JAXA) in Sagamihara and at the National Astronomical Observatory of Japan~(NAOJ) in Mizusawa and Mitaka, Japan. In Section~\ref{section:concept}, we describe the concept of the test flight model as a balloon-borne VLBI station. In Sections~\ref{section:observingsystems} and \ref{section:bus}, the configuration and development of the observing system and gondola bus system are described. In Section~\ref{section:discussion}, we briefly discuss the experiment.  Finally, the conclusion is presented in Section~\ref{section:conclusion}.

\begin{figure*}
\includegraphics[width=\linewidth]{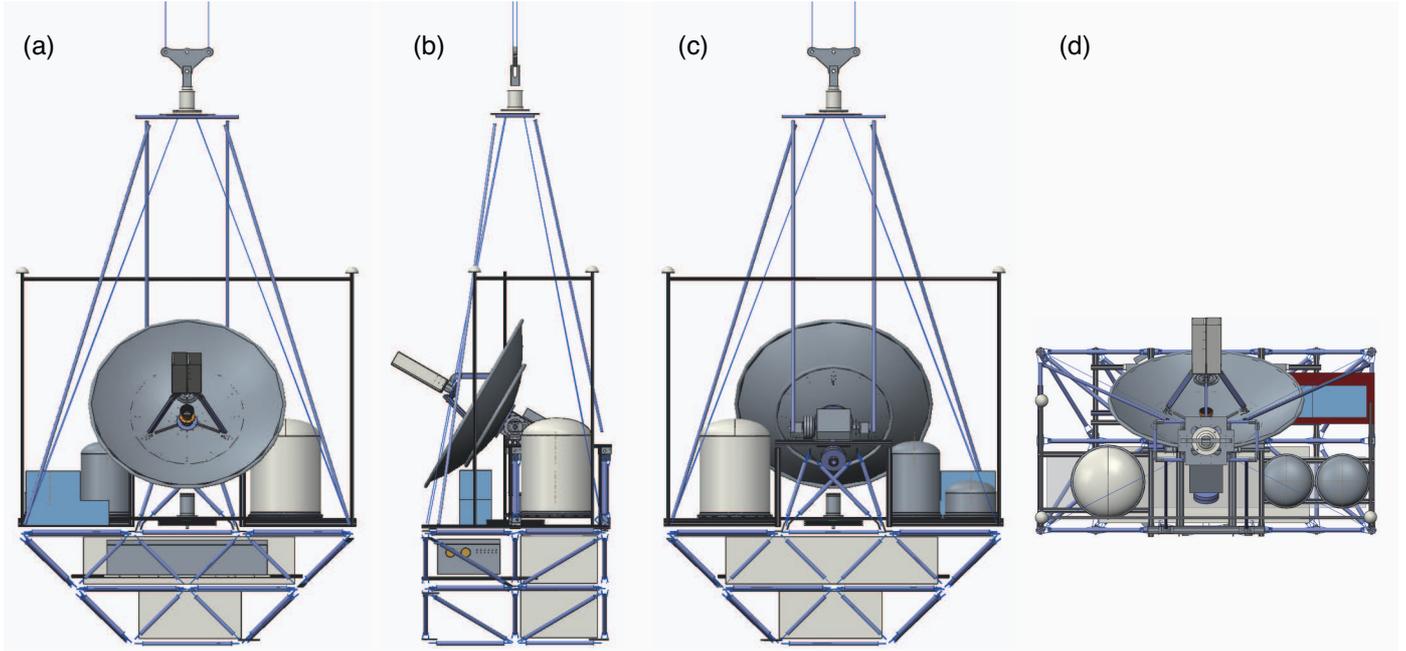}
\caption{Four side views of the balloon-borne VLBI station. Sun shields and ballast boxes are not described in this model. (a)~Front view, (b)~Side view, (c)~back view, and (d)~top view. \label{fig:3sided-views}}
\end{figure*}

\section{Concept of experiment}\label{section:concept}
This paper discusses the first attempt to achieve VLBI observation using a stratospheric balloon. The results of this experimental engineering mission is valuable to evaluate the technical feasibility of future VLBI using balloon-borne radio telescopes. The technical challenges are as follows: (1) implementing an on-board frequency standard clock; (2) implementing an on-board VLBI data recorder; (3) controlling the pointing of the telescope; (4) determining the position of the VLBI element under the environmental conditions of the balloon-borne platform.

\begin{figure}
\includegraphics[width=\linewidth]{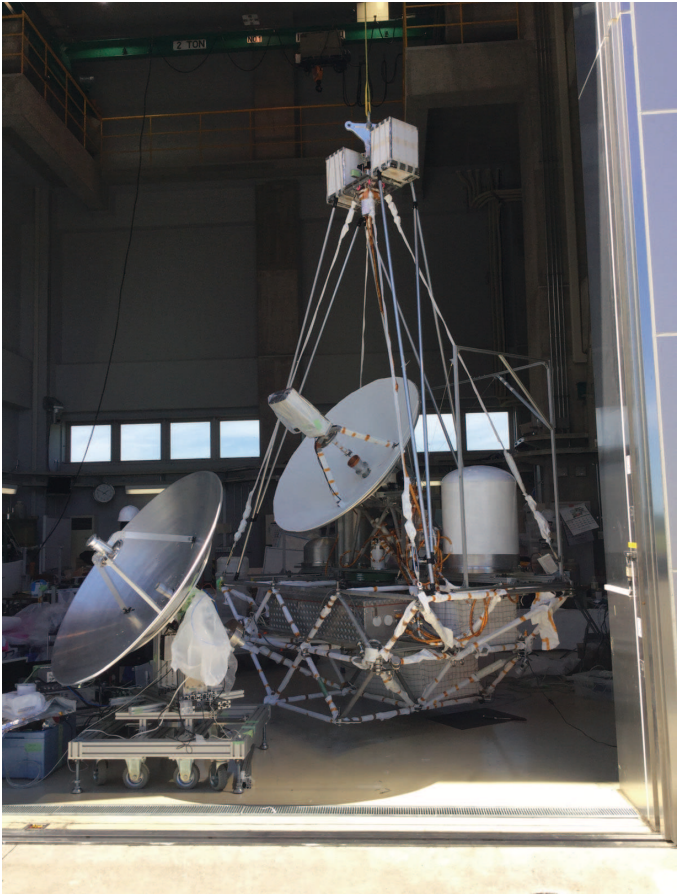}
\caption{Photograph of the balloon-borne VLBI station during a hanging test in the assembly room in JAXA TARF. Sun shields and ballast boxes are not attached to the gondola at this time. The Taiki 1.5-m ground-based station is shown on the left.\label{fig:pictureOnLauncher}}
\end{figure}

\begin{figure}
\includegraphics[width=\linewidth]{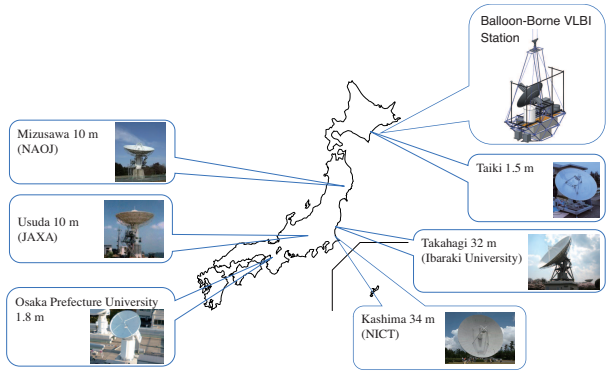}
\caption{VLBI network for test observations in 2017.\label{fig:VLBInet}}
\end{figure}

As the first step, a single-element balloon-borne VLBI station with a radio telescope and K-band ($\sim20$~GHz, $\lambda1.5$~cm) receivers are employed (four side views and a photograph of the station are shown in Figures~\ref{fig:3sided-views} and \ref{fig:pictureOnLauncher}, respectively). Several ground-based VLBI stations distributed over the Japanese islands are incorporated as counterparts of VLBI elements to form baselines ranging from $\sim50$ to $\sim1,000$~km in length during the flight.  
The ground-based radio telescopes used in test observations in 2017 were the Mizusawa~10~m \citep{Kameya:1996,Shibata:1994} operated by NAOJ, the Takahagi 32~m \citep{Yonekura:2016} operated by the Center for Astronomy, Ibaraki University, the Kashima 34~m \citep{Sekido:2013} operated by the National Institute of Information and Communications Technology~(NICT), the Usuda 10~m operated by JAXA, and the Osaka Prefecture University 1.8~m telescope (Figure~\ref{fig:VLBInet}).
As a primary target, we utilize an artificial signal source of the geostationary satellite ``IPSTAR'' (or ``THAICOM-4''), which appears across a frequency range of $\sim19.6$--$20.2$~GHz and includes a beacon at 20.199827~GHz. The IPSTAR signal is very strong and allows us to perform measurements at a high signal-to-noise ratio (SNR), thus it is useful to evaluate of the phase stability of the on-board VLBI components. 
Continuum emission and/or H$_\mathrm{2}$O maser line emission from astronomical objects may be observed as secondary targets if extra time is available at the level flight.

The balloon-borne VLBI station is scheduled to be launched from the Taiki Aerospace Research Field \citep[TARF; ][]{Fuke:2010}, a scientific balloon facility operated by JAXA in Hokkaido, Japan.  A scientific B30 balloon ($30,000$~m$^{3}$), fabricated by Fujikura Co., Ltd., will be used. After ascending for 0.5~h and drifting eastward for 1.5~hours at an altitude of $\sim12$~km in a so-called boomerang flight operation \citep{Nishimura:1981}, the balloon will reach a level altitude of 26~km, at which it will remain for 1.5~h or more for experimental VLBI observations. The total platform weight, not including a 180~kg ballast, is approximately 600~kg. The specifications of the observing and bus systems are listed in Table~\ref{table1}.

\section{Observing Systems}\label{section:observingsystems}

\begin{figure*}
\includegraphics[width=\linewidth]{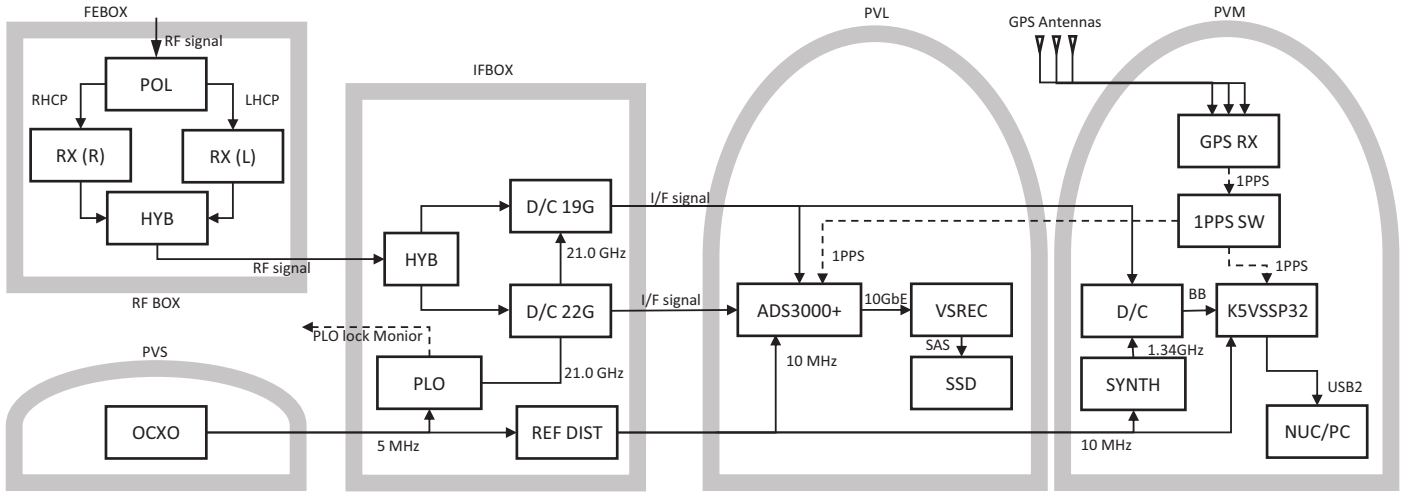}
\caption{Block diagram of the observing systems for balloon-borne VLBI station. Abbreviations in the figure are as follows. POL: polarizer, RX: Receiver for low-noise amplifiers, HYB: hybrid coupler, D/C: downconverter, REF DIST: reference signal distributor, SYNTH: synthesizer, SW: switch, BB: base band signals. \label{fig:blockdiagram}}
\end{figure*}

A block diagram of the on-board observing and bus systems is shown in Figure~\ref{fig:blockdiagram}.

\subsection{Antenna}\label{section:antenna}
The antenna is a Cassegrain-focal design with primary and secondary mirror diameters of $1,500$ and $200$~mm, respectively, and a primary focal length of $488.7$~mm. These optics parameters have been determined to utilize an existing fabrication design of the primary mirror.  
To enhance its rigidity, the parabolic primary mirror is composed of two 3-mm thick layers of aluminum (A1100), which are $1,500$ and $455$~mm in diameter, respectively. The two parabolas were formed through a cold working process using a spinning machine manufactured by Kitajima Shibori Seisakusho Co., Ltd.
Based on a structural modeling analysis, the maximum expected self-weight deformation is $0.2$~mm at a point on the outer edge of the primary mirror.  The surface accuracy of the primary reflector is $\sim0.25$~mm~r.m.s.\ including a self-weight deformation at the elevation of $10$~deg, which was measured by photogrametry using V-STARS together with PRO-SPOT Target Projector\footnote{Geodetic Systems, Inc.} to an accuracy of $0.03$~mm. Therefore, the surface efficiency is expected to be $95.7$\% at $20$~GHz. A subreflector unit with a tripod stay structure was attached during the measurement.
The secondary mirror is $200$~mm in diameter and is fabricated from aluminum A5052 via a machining process. Through laser scanning measurement, the secondary mirror surface accuracy was measured to be $30\ \mu$m~r.m.s.
A corrugated horn has half-angle of $11.7$~deg and illuminates the main reflector with an edge level of $-10$~dB. The optically designed aperture efficiency is $0.59$, and the spillover loss is estimated to be $0.06$. The blocking efficiency of the subreflector unit is $\sim0.05$, thus resulting in a calculated effective radio telescope aperture efficiency of $0.50$.
The primary beam size at the frequency ranges to be used for observations of IPSTAR and the astronomical radio sources is $\sim0.6$~deg.
In order to reduce thermal deformation due to solar heat input, the front and rear sides of the primary mirror were painted white.

\subsection{Frontend: Receivers}\label{section:receivingsystem}
The front end~(FE) systems are contained within an aluminum box, called ``FEBOX,'' that mechanically supports the antenna as a load path using a rotating elevation axis.
The dual circular K-band polarization receiving system comprises a septum-type polarizer \citep{Kaiden:2009}, a Weinreb's low-noise amplifier~(LNA) manufactured by the Caltech Microwave Research Group, a MITEQ LNA for left-handed circular polarization~(LHCP), and a B\&Z LNA and a MITEQ LNA for right-handed circular polarization~(RHCP).
The LHCP and RHCP signals are joined in a combiner and are subsequently transferred together at a radio frequency (RF) regime to the exterior of FEBOX.  
The system minimizes the amount of mechanical twisting torque by employing a single co-axial RF cable passing through the elevation axis via a rotary joint (Section~\ref{section:ACS}).
Single circular polarization observations are made via the alternate use of LHCP/RHCP by depowering the respective LNAs. Only LHCP observation is planned for nominal use in matching the ground-based station receivers as VLBI counterparts, thus relegating the RHCP receiver to the status of a standby redundant system. Note that the IPSTAR signal comes in a linear polarization, while most astronomical radio sources are almost perfectly unpolarized.  Each receiver has an efficiency of $1/2$ toward the IPSTAR signal and astronomical radio sources. 
The system LHCP noise temperature has been measured to be approximately 320~K. The assumed antenna aperture efficiency of 0.50 (Section~\ref{section:antenna}) corresponds to an expected system equivalent flux density~(SEFD) of $\sim 1\times 10^{6}$~jansky~(Jy) for the telescope; the SEFD in RHCP is slightly worse than that in LHCP owing to the receiver's noise temperature differences between the respective LNAs.  The fringe detection sensitivities are presented in Section~\ref{section:VLBIsystem}.  


\subsection{Intermediate frequency Division}\label{section:IF}

The platform's intermediate frequency~(IF) systems are contained in an aluminum box, called ``IFBOX,'' located on the exposed section of the gondola base frame. The RF signal is split through a divider into two bandpass filters, with down-conversions to IF signals made using a first local~(LO) reference signal at $21.0$~GHz. A phase-locked dielectric resonator oscillator~(hereafter referred to as PLO) manufactured by the Nexyn Corporation generates the first LO in synchronization with a 5~MHz reference signal.
Following amplification and transmission from IFBOX, the two IF signals are supplied to respective wide- and narrow-band data recording systems (Section~\ref{section:VLBIsystem}).

IFBOX also contains a distributor that supplies $5$ and $10$~MHz reference signals using a $5$~MHz reference from the oven-controlled crystal oscillator~(OCXO) in a small pressurized vessel~(PVS; Section~\ref{section:VLBIsystem}) located on the outside of IFBOX.
IFBOX is entirely covered with polystyrene foam to reduce the influence of sharp temperature changes in the outside environment. A chamber test of IFBOX in which the stratospheric environment was simulated in terms of ambient temperature and pressure was conducted to confirm the functionality of the filters, amplifiers, and PLO.

The IF signal used for wide-band observation is transmitted to a large pressurized vessel~(PVL) and divided and filtered into four baseband channels~(BBCs) in the ranges $1,024$--$1,536$, $512$--$1,024$, $1,024$--$1,536$, and $1,536$--$2,048$~MHz, corresponding, respectively, to RFs of: 
$19.464$--$19.976$~GHz, 
$19.976$--$20.488$~GHz, 
$22.024$--$22.536$~GHz, and 
$22.536$--$23.048$~GHz. 
The IPSTAR observation is performed using the first two BBCs, while astronomical observation is performed using all four BBCs.
The IF signal for the narrow band of observation is sent to a mid-sized pressurized vessel~(PVM) and down-converted using the second LO reference signal to a 0--32~MHz BBC, corresponding to an RF of 19.660--19.692~GHz for its use in IPSTAR observation.
As the second LO generator, a Pilulu833 phase-locked loop~(PLL) frequency synthesizer module\footnote{Denshiken Co., Ltd.} using an HMC833\footnote{Hittite Microwave Corporation} integrated circuit~(IC) is employed. The second LO is locked in synchronization with a 10~MHz reference signal to generate the 2nd LO frequency 1.34~GHz.

\subsection{Backend: VLBI Systems}\label{section:VLBIsystem}
The frequency standard clock device used for VLBI observations is required to have stability to maintain coherence within unit integration time.  Integration time is necessary to obtain sufficient signal to noise ratio to detect interferometric fringes, which is typically 10~seconds or more for VLBI observations.  The ground-based VLBI stations use a hydrogen maser atomic clock (H-maser) that satisfies this requirement; however, factors, such as weight, physical size, and environmental conditions, dictate the use of different resources for clocks mounted on a flying object.
As a standard frequency generator for the balloon-borne VLBI, we adopted an 8607-B series OCXO, which represents the second generation of OCXOs developed by Oscilloquartz SA using the technique of housing a state-of-the-art Boitier Vieillissement Ameliore~(BVA) stress compensation cut (SC-cut) crystal resonator and its associated oscillator components within double oven technology.
The Allan standard deviation of the frequency stability is lower than $8 \times 10^{-14}$ at $\tau =1$--$30$~sec.     
We conducted a comparison between two OCXO8607s and several atomic clocks based on VLBI experimental observations at 230~GHz. The evaluation results are reported in a separate paper \citep{Kono_BVLBI:submitted}.
In the balloon-borne VLBI platform, the OCXO8607 is installed within the PVS (Section~\ref{section:PV}).

Two VLBI data acquisition systems that can operate simultaneously are employed for the experiment.
The first system is a wide-band VLBI system for observing both IPSTAR and astronomical objects. Four base-band signals (Section~\ref{section:IF}), each with a $512$~MHz bandwidth, are sampled using an ADS3000+ analog-to-digital sampler developed by the National Institute of Information and Communications Technology~(NICT). The sampling is initially performed in 8-bit quantization, and the data are subsequently resampled in a 2-bit quantization at the Nyquist sampling rate after level adjustment, thus resulting in an aggregated bit rate of $8.192$~Gbps. 
The digitized data are recorded at $8.192$~Gbps using a mission-dedicated VLBI software recorder that was developed by upgrading a VSREC VLBI recorder that is being developed by NAOJ \citep{Oyama:2012}. The improvements involve replacing the VSREC hard disk drives with solid-state drives (SSDs) of 4~TB for data and 0.5~TB for operating system and reducing the energy consumption rate by introducing a new motherboard (Skylake\footnote{Intel Corporation.}) and CPU. Hereafter, we refer to this dedicated VSREC as ``VSREC.''
The wide-band VLBI system is installed within the PVL.  
With the 8~Gbps recording rate, the 7-sigma fringe detection sensitivities\footnote{The theoretical baseline sensitivity (1-sigma) is defined as $\eta^{-1} (SEFD_i SEFD_j / (2 B \tau))^{0.5}$ in Jy, where $\eta(\approx 0.8)$ is a correlation efficiency, $B$ is the bandwidth in Hz, and $\tau$ is the integration time in sec.} of the baseline between the balloon-borne station and Takahagi~32~m ground-based telescope is expected to be $\sim3$--$10$~Jy ($T_{\rm sys, 32m} = 40$--$500$~K; \citealt{Yonekura:2016}).  We assume the integration of 1~sec.  In our plan for the flight experiment, 3C~454.3 or 3C~84 ($\sim20$~Jy) is observed as a target of the astronomical object.

The other VLBI system, which operates within the PVM, is a narrow-band system that employs the a K5/VSSP32 VLBI sampler, which is developed by NICT \citep{Kondo:2006}. A base-band signal with a bandwidth of $32$~MHz is sampled in a 2-bit quantization at the Nyquist sampling rate, thus resulting in an aggregated bit rate of $128$~Mbps. Although using K5/VSSP32 makes $256$~Mbps in two IFs available, only one IF includes the most significant component in the IPSTAR signal at an RF of $19.675$~GHz. The digitized data are recorded on the SSD of an Intel$^{\textcircled{R}}$ Next Unit of Computing~(NUC) mini PC.  

To achieve time adjustment, the ADS3000+ and K5/VSSP32 samplers are synchronized to a $10$~MHz reference signal during observation and with a 1~pulse-per-second~(1PPS) signal of the global positioning system~(GPS) before observation.  Linear and quadratic drifts of the OCXO clock will cause time offset to change.  Off-line procedures of fringe search/fitting are applied in correlation/data-reduction processes after the observation.  
       

\subsection{Power detector and spectrometer}\label{section:detectorandspectrometer}

We designed the VLBI element so that it can be also used as a single telescope by adding a power detector software and a spectrometry software. The software processes the digitized data acquired by K5/VSSP32 in which a base-band signal with a width of 32 MHz in semi-real-time with delay of 0.8 seconds every 0.1 second.  The power detector software calculates the square mean. We intend to use the detector output for observations of the continuum emissions of the Moon and Sun.  
The spectrometry software calculates the power spectrum by FFT processing to output the power ratio of the carrier component the IPSTAR signal to the adjacent noisy back ground spectrum. 
At the location of TARF in the eastern part of Hokkaido, IPSTAR has been detected at approximately $2$--$3$~dB using the radio telescope that is to be launched in the experiment.

\begin{table*}
\caption{Specifications of observing and bus systems for balloon-borne VLBI experiment.\label{table1}}
\begin{center}
\begin{tabular}{llr}     
\hline\hline     
Parameters & Unit & Value \\
\hline
Antenna diameter & meter & 1.5 \\
Surface accuracy RMS & mm & 0.25 \\
Aperture efficiency$^{\rm a}$ &  & 0.50 \\
System noise temperature $T_\mathrm{sys}$ & Kelvin & 320 \\
System equivalent flux density (SEFD)$^{\rm b}$  & Jy & 1,006,000 \\
Observing frequency (mode A) & GHz & 19.660--19.692 \\
Observing frequency (mode B) & GHz & 19.464--20.488 \\
Observing frequency (mode C) & GHz & 19.464--20.488 \\
   & &  22.024--23.048 \\
Half-power beam width (HPBW)$^{\rm c}$ & degree & 0.64 \\
1st local oscillator frequency & GHz & 21.000 \\
\hline
Analog-to-digital conversion quantization & bit & 2 \\
Recording rate (mode A) & Gbps & 0.128 \\
Recording rate (mode B) & Gbps & 4.096 \\
Recording rate (mode C) & Gbps & 8.192 \\
SSD media capacity & Tbyte & 4 \\
\hline    
Dimension of gondola base frame (W$\times$D$\times$H) & meter & $2.60\times1.40\times4.18$ \\
Total weight (dry) & kg & 608 \\
Ballast weight & kg & 180 \\
Battery capacity & Wh & 5320 \\
Power consumption (typical) & W & 370 \\
\hline        
\end{tabular}
\end{center}
\begin{footnotesize}
$^{\rm a}$Including the optics efficiency, the reflection loss, the blocking factor, and the spill over loss.\\
$^{\rm b}$Including an pointing error of 0.21~degree at maximum.\\
$^{\rm c}$Defined at the center frequency of mode~B.
\end{footnotesize}
\end{table*}

\section{Bus Systems}\label{section:bus}

\subsection{Pressurized Vessels}\label{section:PV}
The ambient pressure and temperature in the stratosphere are expected to be $\sim1/50$~atm and $\sim-50^\circ$C, respectively, during the level flight of the platform. Because many commercial off-the-shelf (COTS) products are used for the experiment, the three pressure vessels~(PVs) discussed in the previous sections are used to maintain environmental conditions similar to those on the ground of $\sim1$~atm and room temperature. All the PVs are made of 4-mm thick domes and 11-mm thick aluminum~(A5052) base plates fabricated by Kitajima Shibori Seisakusho Co., Ltd. The PV domes and base plates are sealed together by tightening them with a mechanical clamp, and the base plates are equipped with several hermetic connectors for power supply lines, signal lines and co-axial cables. Two air cocks are also attached on each base plate for gas replacement of air potentially containing moisture by dry nitrogen before the launch.

The small pressurized vessel~(PVS) has a diameter of 400~mm and a height of 271~mm. It contains the components that require temperature stability, including the OCXO, the three-axis gyroscopes, and the three-axis accelerometers. Much of the internal volume of the PVS is filled with a thermal insulating material to prevent the devices from suffering rapid changes of temperature. The exterior is an unpainted aluminum surface that is expected to be insensitive to ambient radiation fields.

The PVM has a diameter of 400~mm and a height of 611~mm and contains several Raspberry~Pi operational PCs (OPCs)\footnote{Raspberry Pi Foundation.} for command/telemetry operation, attitude determination (Section~\ref{section:ADS}), attitude control (Section~\ref{section:ACS}), and housekeeping. In addition, the PVM houses the GPS compass receiver, the downconverter, a 1PPS distributor, one of the two VLBI systems (K5/VSSP32; Section~\ref{section:VLBIsystem}), three Ethernet switching hubs, and relay array modules to power on/off controls and reboot dysfunctional devices.  

The PVL has a diameter of 550~mm and a height of 788~mm and contains the wide-band VLBI system that comprises the ADS3000+ and VSREC (Section~\ref{section:VLBIsystem}). The following two conflicting requirements make the thermal design of this unit difficult. The considerable power consumption ($\sim280$~W) of these two instruments results in a high rate of heat generation during their recording operation. On the other hand, the retention of startup temperature condition is required during standby operation to obtain power saving ($\sim40$~W) before VLBI observations. Owing to this dilemma in thermal design, we decided to place actively controlled fans between a heat dissipating compartment and a thermally insulated compartment, resulting in a dual-compartment PVL.
The exterior surface of the PVL dome is painted white to aid in radiation cooling.
The assembly performed quite well in a thermal vacuum test in which an environment equivalent to that during a level flight was simulated. The air temperature of the thermally insulated compartment containing the VLBI instruments did not decrease below $\sim0^\circ$C during the standby mode. As the instruments went into full operation without fan activation, the air temperature eventually increased to a hazardous level ($>40^\circ$C); however, the air temperature maintained at $\sim+15^\circ$C during fan activation, which was expected based on our analytical thermal design.

\subsection{Attitude Determination Systems}\label{section:ADS}
The attitude determination system~(ADS) provides operational attitude determination (OPC-AD) under the control of a Raspberry Pi~3 model~B\footnote{Developed by Raspberry Pi Foundation.} installed within the PVM. The function of the OPC-AD is to collect information from the attitude sensors, calculate an attitude determination solution, and subsequently send attitude control commands to the operational PC for the attitude control system (OPC-AC; Section~\ref{section:ACS}).

The attitude determination solutions in the azimuth (AZ) angle are calculated by combining the outputs of a coarse sensor and a fine sensor through a complementary filter. We use a JG-35F fiber optic gyro (FOG)\footnote{Japan Aviation Electronics Industry, Ltd.} as the fine AZ-axis sensor. The FOG has a resolution of a few arcseconds, and a coarse sensor that is less accurate but provides an absolute angle is used to compensate for apparent velocity drift in the gyro output. OPC-AD can select an on-board GPS compass, a geomagnetometer, or a solar angle sensor as the input to the complementary filter.
The GPS compass uses a PolaRx2e@ GPS receiver\footnote{Septentrio Satellite Navigation N.V.} coupled with one L1/L2 antenna and two L1 antennas\footnote{Sensor Systems Inc.}. This compass configuration has been previously used in flight in the pGAPS experiment at TARF \citep{Fuke:2014}. A three-axis MAG-03~MS magnetic sensor\footnote{Bartington Instruments Ltd.} is used as the geomagnetometer; previously, it was used in flight and subsequently retrieved from the sea in an ISAS balloon-borne infrared astronomy mission. The sun sensor is an imaging camera based on a Raspberry Pi camera board with a field of view of approximately $60\times50$~deg.
The resulting output from the complementary filter has several arcseconds of stability and an absolute accuracy of $\sim0.1$--$1$~deg.

The attitude determination solutions in the elevation (EL) angle are also calculated using a complementary filter. As a fine sensor along the EL axis, CRH02-025 silicon ring gyroscope sensors\footnote{Silicon Sensing Systems Ltd.} based on micro-electro-mechanical system technology are mounted on the radio telescope's frame. The EL gyro has a resolution of a few arcseconds and the pitch output from the three-axis geomagnetometer is used as one of the inputs of the complementary filter to compensate for the apparent velocity drift of the gyro output. Note that the geomagnetometer is located on the gondola base frame, and the difference in coordinate system between the telescope and gondola frames is compensated by the optical encoder used for driving the EL drive (Section~\ref{section:ACS}). In other words, the attitude determination in EL is not affected by the pitch component in the pendulum motion between the balloon and the gondola as both the gyroscope and the geomagnetometer are influenced by the inertial space system. Thus, the output from the complementary EL filter also has several arcseconds of stability and an absolute accuracy of $\sim0.1$--$1$~deg.

As the primary beam size of the radio telescope is approximately 0.5~deg at FWHM (Section~\ref{section:antenna}), an absolute angle determination accuracy of $0.1$~deg angle is required for pointing to the target. As solely the output of the compensation filter does not satisfy this requirement, our operational strategy involves performing a raster scan on the sky centered on the predicted location of IPSTAR and pointing to it using the output of the spectrometer (Section~\ref{section:detectorandspectrometer}). Following this, we calibrate the coarse sensors for subsequent observations of astronomical objects.  Astronomical maser objects normally used in ground-based VLBI telescopes can not be used for pointing calibration due to insufficient sensitivity of the on-board radio telescope.

We are planning to use star trackers for fine attitude determination in future missions.  In this experiment, two test models of the star tracker are mounted parallel to the optical axis of the radio telescope to the back of the secondary mirror.  We will make shooting tests in the stratosphere with these test models, although these are not used for on-flight attitude determination in this experiment.  The star trackers in the daytime stratosphere requests designed to prevent saturation due to foreground radiation from the sky.   
Based on the conceptual design of DayStar \citep{Truesdale:2013}, we fabricated the following two daytime star trackers to be used as cameras to detect star images: UI-3370CP-NIR-GL\footnote{IDS Imaging Development Systems GmbH.} using a CMOSIS\footnote{ams AG.} image sensor and UI-1240ML-NIR-GL\footnote{IDS Imaging Development Systems GmbH.} using an E2V\footnote{Teledyne e2v (UK) Ltd.} image sensor with an LP610 red longpass filter\footnote{Midwest Optical Systems, Inc.} with a cut-on wavelength of $\lambda 610$~nm to reduce the blue sky foreground. The trackers are equipped with baffles; to enable an obstruction light avoidance of 12~degrees, they have lengths of approximately 400~mm and square apertures of roughly 120~mm.  In the stratosphere, we perform a star shooting test at various sun angles and altitudes including ascent.  In order to evaluate the design concepts of the baffle and camera system, we measure the sun angle dependence of stray light and the relationship between the star magnitude and SNR, respectively.

The pitch and roll orientations of the gondola base frame are monitored by CRH01-025 and CRS39-02 silicon ring gyroscope sensors\footnote{Silicon Sensing Systems Ltd.}, an LCF-23102-D tilt meter, an LCF-25302-D accelerometer\footnote{Jewell Instruments LLC.}, and a MMA84513-D accelerometer, all of which are installed in the PVS (Section~\ref{section:PV}). Orientation is achieved with respect to the following two frames: an pendulum orientation defined with respect to the gravity vector, and a static orientation defined with respect to the flight train direction based on the imbalance in the mass distribution following ballast discharge.  Gyro sensors respond the former component, while accelerometers sense the latter one.   If static orientations with respect to the AZ gyroscope's axis remain in pitch and roll, a fraction of the pendulum motion would contaminate the output from the AZ gyroscope; therefore, a determination of the dynamical orientation by the OPC-AD is necessary for real-time correction of the AZ determination by taking static orientation into account.

\subsection{Attitude Control Systems (ACS)}\label{section:ACS}
The radio telescope is driven on an azimuth--elevation (AZ--EL) mount coordination system. Attitude control along the AZ axis is achieved by yawing the entire gondola using a reaction wheel~(RW) equipped at the center of the gondola and is aided by the coarse ``PIVOT'' AZ actuator located on top of the gondola. Pointing along the EL is performed by rotating only the radio telescope frame.

All of the actuators have direct drive motors, which are superior to mechanical geared motors as they (1) experience no backlash, (2) are unbreakable even under an unexpectedly large external torque, and (3) provide quasi-free rotation on demand by depowering to avoid the influence of external disturbance torques. However, the direct drive motors are both dimensionally larger and heavier than geared motors. Kollmorgen's frameless brushless direct drive motors are applied to the RW (Model KBMS17H03-D), PIVOT, and EL drives (KBMS43S02-B). Sixteen-bit incremental shaft encoders are used for RW and PIVOT (DFS60A-THAK65536\footnote{SICK AG.}), while a non-contact optical encoder read-head and $\phi150$-mm stainless steel ring with $20\ \mathrm{\mu m}$ pitch graduations combination (Model Ti2601 and RESM20USA150\footnote{Renishaw plc.}) is used for the EL drive. The motors are driven using digital servo drivers (``Whitsle\footnote{Elmo Motion Control Ltd.}'') controlled from the OPC-AC via RS232C serial communication at a rate of 10~Hz.

The PIVOT drive plays the following two roles: unloading over-accumulated momentum of the wheel and canceling the twisting torque that occurs when the flight train connects to the balloon, which can rotate randomly at a typical rate of $\sim0.1$~rpm going up to $\sim1$~rpm. By turning off PIVOT's power until it needs to be activated for unloading of accumulated RW momentum, the gondola's yawing should be practically free from the perturbation torque generated by the balloon. A ground-based test resulted in a controlled stability of $\sim0.01$~deg in the AZ control during the simultaneous activation of both the RW and PIVOT drives.

Pointing control along the EL is performed by rotating the radio telescope frame using the elevation actuator, which was designed on the basis of the mechanical concept of the Wallops ArcSecond Pointing (WASP) system \citep{Stuchlik:2014,Stuchlik:2015}. The WASP concept involves the use of a shaft that is continuously rotating at a constant rate of $\sim30$~rpm, with both the gondola base frame and the telescope frame floating on the rotating shaft via axial roll bearings. As a result of this configuration, no static friction should affect the telescope with respect to the gondola during the switch-backing of the pendulum, and the dynamic friction on the axial roll bearings on the left and right sides of the EL axis should somewhat balance each other.  The radio telescope is balanced with respect to the EL axis by a counter weight. In principle, the telescope should remain fixed with respect to the inertial space coordinates even if a disturbance, such as pendulum vibration is added to the gondola frame. Accordingly, the EL drive motor is activated with a very slight force when a pointing displacement occurs. A ground-based test resulted in a controlled stability of $\sim0.01$~deg in the EL control during the pendulum motion of $\sim1$~deg or less, and we found no signature of potentially increased pendulum motion owing to feedback control excitation. Note that at $EL \neq 0$~deg, it is necessary to utilize two-axis gimbals to perform rolling along the EL to compensate for the drift in the antenna pointing arising from the pendulum motion in the roll direction in practice \citep[e.g.,][]{Ward:2003}. For the initial flight test, the WASP-like mechanism in the gondola system has a single one-axis actuator along the EL.

The performance of the attitude controls under hanging testing will be reported in a separate paper.  

\subsection{Position Determination}\label{section:positiondetermination}
Fluctuation in the line-of-sight direction component of the position of the VLBI station degrades the coherence of the received radio wave.  The fluctuation must be kept below about $1/20$ wavelength to maintain the coherence.  This is the requirement for signal integration during the time needed to improve the $SNR$ (typically 10~s or longer) for the fringe detections of astronomical radio sources.  Our approach is to estimate the position change using onboard sensors and correct the received radio wave by rotating the phase during correlation processing.  

We install an accelerometer in the radio telescope frame along the telescope's optical axis to estimate the position change successively by integrating the second-order output of the accelerometer.    To do this, we use a JA-40GA02 accelerometer\footnote{Japan Aviation Electronics Industry, Ltd.}, which has a sensitivity of up to 0.7~$\mu$G. Based on a numerical simulation assuming combined translational and pendulum motion involving shear disturbances of wind speed, we expect a sequential position determination accuracy $\sim40$~$\mu$m for 10~$\mu$G measurement noise. Although the output of the accelerometer usually includes systematic errors in offset and drift, these can be solved through fringe search processing of the interferometer. The output of the accelerometer is recorded along with the time stamp during the VLBI observation and subsequently analyzed following observation and recovery.

Another position determination method is to estimate the pendulum angle of the balloon-gondola system by integrating the velocity output of the pitch component from the gyroscope mounted on the gondola base frame and multiplying the result by the physical length of the pendulum to obtain the change in position. Application of this strategy in a VLBI observation using the flight model under pendulum disturbance during hanging test produced a detectable clear fringe; the results to be reported in a separate paper \citep{Kono_BVLBI:submitted}.

\subsubsection{Power Supply Unit}\label{section:PSU}
As a power supply, the platform employs PC40138LFP-15Ah lithium-ion battery cells\footnote{Phoenix Silicon International Co.} (3.3~V, 15~Ah) using LiFePO4 as the positive electrode material.     
Each battery module comprises eight series-connected cells, assembled by Wings Co.\ Ltd, producing 26.4~V and 396~Wh/module. The gondola is equipped with 14~modules connected in parallel, producing approximately 5,500~Wh in total. The typical power consumption during level flight is expected to be 400~W, reaching a maximum of approximately 600~W.         
The PC40138LFP-15Ah battery cell has a recommended operational temperature range of $-10$--$+55^\circ$C during discharge. We conducted discharge tests at ambient temperatures of $-20$, $0$, and $+25^\circ$C and found that approximately $50$, $75$, and $100$\% of the total capacity, respectively, were available under these conditions. 

The battery modules and housekeeping electronics are contained within a W$1,250\times$D$480\times$H$270$~mm$^{3}$ aluminum box manufactured by Shin Kowa Industry Co., Ltd\footnote{http://www.shin-kowa.co.jp}.  The box is made doubly waterproof in the case it gets submerged in seawater to recover the gondola system through the inclusion of a thermal insulation box fabricated from expanded polystyrene foam with thermostatically controlled heaters.  Vent filters\footnote{GORE$^{\textcircled{R}}$ VENT; https://www.gore.com/products/categories/venting} located on the boxes are used to adjust the internal pressure to conform to altitude-related changes. A thermal vacuum test conducted in a chamber-recreated stratospheric environment resulted in an internal temperature of $\sim10^\circ\mathrm{C}$, which is consistent with the predictions of our numerical thermal analysis.

\subsubsection{Gondola Structure}\label{section:gondolastructure}
The gondola base frame was designed to meet requirements in terms of weight, strength (including low-temperature environments), nonmagnetism, the field of view of the telescope, and cost. The solution we obtained was a boat-shaped truss structure formed from SUS304 pipe rods, which helped to increase the specific rigidity of the gondola base frame. As our initial survey indicated that no commercially available truss system could satisfy our requirements from the perspectives of weight and cost, we used Technotrass$^{\textcircled{c}}$ building structural parts\footnote{http://www.ts.org/techno\_truss.html} developed by TechnoSystems. Twelve varieties of pipe rods with diameters of 34~mm and thicknesses of 1~mm were specially manufactured by Gantan Beauty Industry Co., Ltd\footnote{https://www.gantan.co.jp/} and assembled. To satisfy the 10~G static loading condition imposed on gondola systems launched at TARF, we used SUS304, which neither exhibits low temperature brittleness nor affects the geomagnetic sensor. We validated the structural design through structural analysis including a buckling phenomenon, resulting in a unit meeting the 2,600~mm (width) $\times$ 1,400~mm (depth) $\times$ 840~mm (height) dimensional requirement and the 100~kg mass requirement.

The gondola width was designed not to hinder the aperture of the radio telescope with its wire cables from four corner foot points (Figure~\ref{fig:3sided-views}(a)). The four wire cables are connected to the flight train via PIVOT, which blocks the aperture of the radio telescope from an elevation angle of $60$~deg or above, making this the practical elevation limit of the telescope. The pressurized vessels, battery box, and radio telescope are supported by aluminum frames and are further mechanically interfaced to the nodes of the trusses. As the linear thermal expansion coefficients of SUS304 and aluminum differ, the truss structure and aluminum frame are fastened with a degree of freedom left to accommodate the expected direction of thermal deformation. Because the entire gondola will be recovered from the sea at the end of the flight mission, a float fabricated from expanded polystyrene foam ($550\ell$ in volume) is enclosed within the truss structure, with an additional $90\ell$ expanded polystyrene float installed near PIVOT. Combined with the volume of the pressurized container and the battery box, the net buoyancy is equivalent to approximately $1,000\ell$. We analyzed the expected altitude after landing with respect to the mass and buoyancy distributions using the {\tt Grasshopper}\footnote{Grasshopper$^\mathrm{TM}$ graphical algorithm editor with integrated 3D modeling tools.} and found that PIVOT, which is the point that will be grappled by the crane of the recovery ship, should float on the sea surface. The overall gondola structure will either float with the antenna facing downward (with a $70\%$ probability) or with it facing upward ($30\%$ probability). Four ballast boxes are installed at both lower sides of the gondola base frame, and approximately 180~kg of iron powder will be loaded as ballast.

\section{Discussion}\label{section:discussion}
The balloon-borne VLBI station described in this paper has been fully developed. A first launch attempt from TARF was scheduled on July 24, 2017, but it was canceled owing to changes in the wind condition at ground level. Through technical tests conducted at TARF before and after the event, interferometric fringes were successfully detected in all the baselines of the balloon-borne VLBI station (under pendulum disturbance) toward the ground-based VLBI stations \citep{Kono_BVLBI:submitted}.  
The rescheduled single element flight will be a technical feasibility demonstration of balloon--ground baselines at $\sim20$~GHz.  A future goal will be to establish balloon--ground baselines at higher frequencies or balloon--balloon baselines via simultaneous balloon flights, providing a stratospheric VLBI imaging array free from atmospheric effects.

\begin{figure*}
\includegraphics[width=\linewidth]{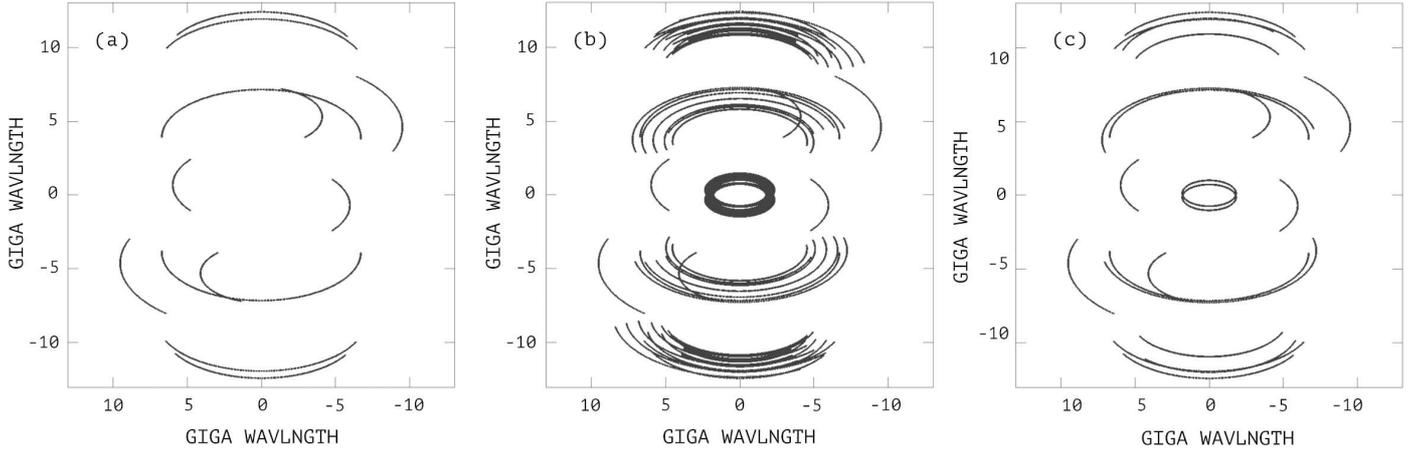}
\caption{Simulations of $uv$-coverages assuming seven-day flights around Antarctica and observation toward the Galactic center. The observing frequency is 350~GHz. (a)~Four ground-based stations. (b)~Four ground-based stations and one balloon-borne station. (c)~Four ground-based stations and one additional Antarctic ground-based station.\label{fig:uv}}
\end{figure*}

Next, we discuss the advantages and future prospects of balloon-borne VLBI. The system configuration characteristics relevant to the current experiment are those of the devices mounted on-board, namely, the frequency standard clock and the data recorder. On the other hand, the VSOP satellite had no frequency standard clock or a data recorder.  Link systems provided reference signal on uplink and data transfer on downlink.   This system configuration was complicated and costly.  As a result, the observation was limited within the range of operating time of ground-based tracking stations.  Less severe environmental conditions for scientific balloon payloads make it easier to mount onboard these critical components for VLBI, which brings about the advantage of simplifying the system configuration for a VLBI station.    The ability to recover the recording media is another advantage of balloon-borne missions over conventional space missions. Given the progress that has been made in increasing recording media densities, a recording rate of 10--100~Gbps could potentially be applied to enhance the sensitivity of each VLBI element (e.g., \citealt{Whitney:2011,Oyama:2016,Kim:2018}).  However, such a high data rate transmission on a radio link is unrealistic (for comparison, the {\it VSOP} and {\it RadioAstron} satellites achieve rates of 0.128~Gbps).  Thus, to achieve next generation space VLBI, further innovations such as optical communication are required.  Such sensitivity improvement is crucial to balloon-borne/space VLBI because the physical diameter of an on-board radio telescope within the payload must be limited to a few meters or less at submillimeter/THz regimes.  In the case of a 2-m diameter for a balloon-borne VLBI telescope ($SEFD = 130,000$~Jy, an aperture efficiency of 63\%, $T_{\rm sys}=100$~K), the 7-sigma fringe detection sensitivities at 350~GHz between the balloon--ground baselines are expected to be $\sim400$~mJy and 50~mJy toward the 12-m ALMA telescope ($SEFD = 6000$~Jy) and phased-up ALMA ($SEFD = 100$~Jy; \citealt{Matthews:2018}), respectively.  We assume here a data recording rate of 64~Gbps (e.g., \citealt{Kim:2018}) and the integration of 10~sec.

\begin{figure*}
\centering
\includegraphics[width=0.66\linewidth]{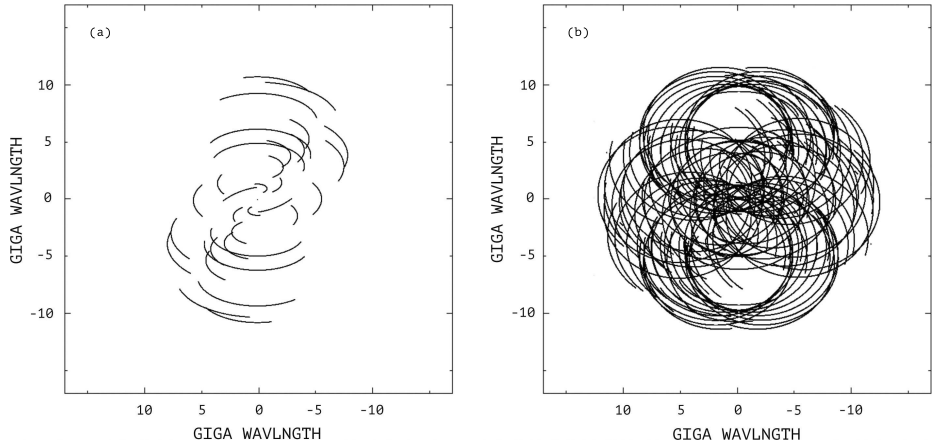}
\caption{Simulations of $uv$-coverages of 24~h space VLBI observation toward the Galactic center. The observing Frequency is 230~GHz. (a)~Eight ground-based stations without space VLBI. (b)~One space VLBI station on a low-Earth orbit (altitude 500~km) and eight ground-based stations.\label{fig:space-uv}}
\end{figure*}

Another advantage of balloon-borne stations is their mobility. One factor that influences image quality is the distribution and density of the sampling space, which is called ``$uv$-coverage''.   The $uv$-coverage represents the assembly of trajectories of baseline vectors seen from an observed celestial target. The number of trajectories is expected to be $N(N - 1)/2$, where $N$ is the number of ground-based stations. Because ground-based VLBI utilizes the rotation of the Earth to generate baseline vector variation, only a series of the identical trajectories can be obtained everyday. However, adding a moving balloon-borne station increases the daily variation of trajectories for $uv$-coverage in the following manner:
\begin{equation}
\frac{N(N - 1)}{2} + \left( \frac{M(M - 1)}{2} + NM \right) \times D, \label{eq:baselines}
\end{equation} 
where $M$ is the number of moving stations and $D$ is the number of days. Figure~\ref{fig:uv} shows the results of a $uv$-coverage simulation for a seven-day observation. Due to the atmospheric phase fluctuations, there may not be so many ground stations that can practically make VLBI observations at the 350-GHz band (e.g., the James Clerk Maxwell Telescope~(JCMT) or Submillimeter Array at Mauna Kea, the South Pole Telescope (SPT), the Large Millimetre Telescope~(LMT), and the phased-up ALMA at Atacama).  The $uv$-coverage made by these four ground-based telescopes is shown in Figure~\ref{fig:uv}~(a). We here consider a long-duration balloon flight over Antarctica, which would typically take two weeks to circumnavigate the South Pole once \citep[e.g.,][]{Seo:2008}. Figure~\ref{fig:uv}~(b) shows the improvements obtained by adding one Antarctic-orbiting balloon-borne station, while Figure~\ref{fig:uv}~(c) shows the improvements obtained by adding one ground-based station to the Antarctica region. Compared to the six baselines achievable with four stations in case (a), in case (b), the number of baselines is increased by 28 (to a total of 34), corresponding to an array of approximately nine ground-based stations.  By contrast, in case (c) the number of baselines has increased by only four (to a total of 10). Thus, adding a circumpolar balloon station can lead to a significant increase in the number of daily baselines. This benefit peculiar to the use of moving stations has been produced by platforms such as the {\it VSOP} and the {\it RadioAstron} satellites. 
Image quality plays an important role in imaging astrophysical phenomena with complex shapes, such as black hole shadows.  A significant increase in the number of baselines could be a decisive factor in the evolution of black hole science through high-fidelity/quality imaging.
  
Based on the effectiveness of the balloon-borne VLBI, it can be expected to develop into a satellite-based VLBI as a future mission \citep{Palumbo:2018}.  Figure~\ref{fig:space-uv} shows $uv$-coverage simulations for a space VLBI at 230~GHz over 24~h generated using the Astronomical Radio Interferometer Simulator \citep[ARIS;][]{Asaki:2007}. Although a different configuration of ground telescopes (potential eight stations; e.g., \citealt{Asada:2017}) is used in this simulation, important improvements are achieved through the use of a low-Earth orbit for the satellite, which at an altitude of 500~km would take only 94~min to orbit the Earth. As this equates to 15 orbits of the Earth over 24~h, which is equivalent to $D = 15$ in Equation~\ref{eq:baselines}, the number of baselines increases significantly for even one observational day theoretically.

\section{Conclusion}\label{section:conclusion}
The balloon-borne VLBI station has been developed for performing radio interferometry in the stratosphere, where the radio telescope is free from atmospheric effects.  The first launch from TARF was canceled owing to changes in the wind condition at ground level.  The rescheduled single element flight will be a technical feasibility demonstration of the balloon-borne VLBI by establishing balloon--ground baselines at $\sim20$~GHz as the first step.
  
This paper described the system design and development of a series of observing instruments and bus systems.  The characteristics of system configuration are that two critical devices for a VLBI station, namely, the frequency standard clock and the data recorder, are mounted on-board.  This configuration maximizes the recording bandwidth and observation time without being limited by the uplink/downlink of the ground tracking station.  Furthermore, the mobility of the balloon-borne VLBI station can improve the spatial sampling coverage by increasing the number of baselines by observation days.  This benefit cannot be obtained with conventional arrays that solely comprise ground-based stations.

\section*{Acknowledgments}
Our deep appreciation goes to Kanaguchi,~M., Miyaji,~Y., Komori,~A., Kobayashi,~H., Ogi,~Y., Takefuji,~K., Tsuboi,~M., and Manabe,~T., Aoki,~T., for their invaluable supports to this project.   
A special note of thanks to all the staff involved in the development and operation of the Taiki Aerospace Research Field~(TARF), a facility of the Japan Aerospace Exploration Agency (JAXA); especially, we received generous support from Yoshida,~T. Saito,~Y., Koyanagi,~K., and Sasaki,~A.  Scientific Ballooning Research and Operation Group and the laboratory of infrared astrophysics offer development infrastructures in the Sagamihara campus of the Institute of Space and Astronautical Science~(ISAS), which is a branch of JAXA.
We are also grateful to the staff and students involved in the development and operation of the Mizusawa VLBI Observatory of the National Astronomical Observatory of Japan~(NAOJ), which is a branch of the National Institutes of Natural Sciences~(NINS). The Mizusawa VLBI Observatory is responsible for the operations of the correlator for the balloon-borne VLBI experiment. A large part of the large pressurized vessel (PVL) has also been developed at the Mizusawa VLBI Observatory: we appreciate the contributions of Matsueda,~C., Asakura,~Y., Matsukawa,~Y., Nagayama,~T., Nakamura,~H., Nishikawa,~T., and Yamada,~R.  
The encouragement from the Japan VLBI Consortium committee were invaluable for continuing the development activities of this experiment.    
Part of this activity is carried out under the collaborative research agreement between RIKEN and JAXA.  
This work was supported by JSPS KAKENHI, Grant Numbers 26120537, 17H02874(AD), 16K05305 (YK), the Casio Science Promotion Foundation (AD), the Inamori Foundation (YK), and the Sasakawa Scientific Research Grant (SN).  
This study was partially supported by JAXA's competitive grants for Strategic Basic Research and Development (HT) and Expense for Basic Development of On-board Instruments and Experiment (YK).

\bigskip
\bigskip


\end{document}